\documentclass[oneside, a4paper, 12pt]{article} %
\usepackage{graphicx}
\usepackage{psfrag}
\usepackage[ english]{babel}
\usepackage{longtable}
\usepackage{cite}
\usepackage{amsmath,amssymb,amsthm}
\usepackage[usenames]{color}

\def\title#1{{\bf\Large #1\par}\vspace{5mm}}
\def\author#1{{#1\par}\vspace{3mm}}
\def\affiliation#1{{\it #1\par}\vspace{2mm}}

\RequirePackage[12pt]{extsizes}
\RequirePackage[left=25mm,right=15mm,top=1.5cm,bottom=20mm,
bindingoffset=0cm]{geometry}

\newcommand{\bs}{\boldsymbol}
\newcommand{\beq}{\begin{equation}}
\newcommand{\eeq}{\end{equation}}

\newcommand{\Om}{\Omega}
\newcommand{\om}{\omega}

\newcommand{\qq}{\qquad}
\newcommand{\q}{\quad}
\newcommand{\lm}{\lambda}

\newcommand{\const}{\rm{const}}

\newtheorem{rem}{Remark}

\makeatletter
\newcommand\tocinarticle{%
     \begin{flushleft}
     \end{flushleft}
\@starttoc{toc}%
   }
\makeatother

\begin{document}
\begin{center}

\title{
Circular  orbits of a ball on a rotating conical
turntable
}

\author{
{Alexey V. Borisov$^{1, 2}$, Tatiana B. Ivanova$^{1, 3}$, Alexander A.
Kilin$^{1, 2}$}
and Ivan S. Mamaev$^{3}$\\
}

\affiliation{
${}^{1}$
 Moscow Institute of Physics and Technology (State University)\\
141700, Russia, Dolgoprudny, Institutskii per. 9 \\[2pt]
${}^{2}$ Udmurt State University\\
426034, Russia, Izhevsk, ul. Universitetskaya 1
 \\[2pt]
${}^{3}$Kalashnikov Izhevsk State Technical University\\
426069, Russia, Izhevsk, ul.Studencheskaya 7
}

\end{center}

\date{\today}

This paper is concerned with the study of the rolling without slipping
of a dynamically symmetric (in particular, homogeneous) heavy
ball on a cone which rotates uniformly about its symmetry axis.
The equations of motion of the system are obtained, partial periodic solutions are found and
their stability is analyzed.


\section{Introduction}

This paper studies the motion of a dynamically symmetric (in particular,
homogeneous) heavy ball rolling without slipping on a cone. The cone
rotates uniformly about its symmetry axis.
To describe the motion, we use
the nonholonomic rolling model, assuming that the motion is subject to
a linear inhomogeneous nonholonomic constraint, which corresponds to
the condition that the velocities of the contacting points
on the surface of the ball and the rotating plane be the same.
There is no rolling resistance.

This problem is interesting not only from a mechanical point of view
(as an example of an integrable system with
nonholonomic constraints), but also because there exist
various analogies of the system under consideration with
systems that are studied in other areas of theoretical physics.
For example, the recent papers~\cite{cone_Gary, cone_English}
 (see also references therein) make an analogy between
the motion of a ball on the surface of a funnel
and the motion of a heavy body in a gravitational field.
Special attention is given to the rolling of a ball without
slipping on a \textit{fixed}
funnel with a variety of  shapes (including the case where
the surface is conical) to search for Keplerian orbits.

In Ref.~\cite{Zengel} an analogy is drawn between the
motion of a ball on the surface of a \textit{rotating} cone and
a charged particle in an electromagnetic field, and use is made of
approximate equations of motion which ignore the time
dependence of the normal to the
surface of rolling at the point of contact and the change in the angular
velocity of rotation of the ball about the normal.

In this work, we present equations of motion of a ball on a
cone taking
into account the time dependence of the normal to the surface
at the point of contact, and discuss the adequacy of
the assumptions made in Ref.~\cite{Zengel}

This paper also gives a detailed stability analysis of
partial solutions which correspond to rolling along circular
trajectories (orbits) at different cone apex angles $\theta\in(0, \pi)$.

\section{Derivation of equations of motion}
\label{sec1}

\begin{figure}[!h]
\centering
\includegraphics[width=0.7\linewidth]{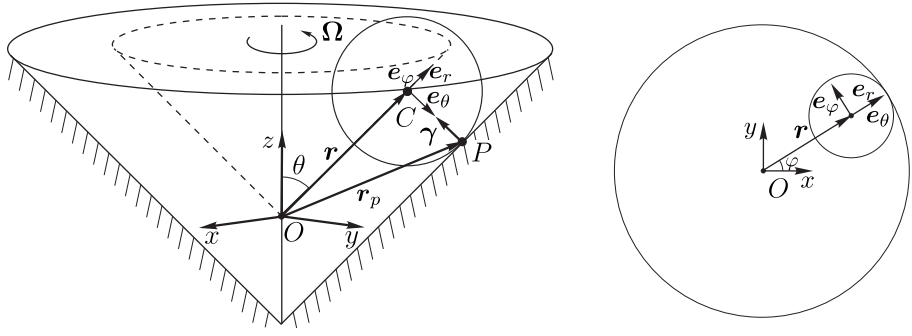}
\caption{A ball on a rotating cone. In the left figure, the unit vector $\bs e_\varphi$
is directed perpendicularly to the figure plane (from the observer).
 The right figure shows a view from above, the~vector~$\bs\Om$ is~directed perpendicularly to the
figure plane (to the observer),
the unit vectors $\bs e_\theta, \bs e_\varphi$ lie in~the~same plane perpendicular to the figure plane.}
\label{comm_01}
\end{figure}

Consider the rolling without slipping of a heavy completely dynamically
symmetric (in particular, homogeneous) ball on the surface of a circular
cone whose symmetry axis is vertical. The cone rotates about the symmetry
axis with constant angular velocity $\bs{\Omega}$ (Fig.~\ref{comm_01}).

Let $\theta\in(0,\pi)$ denote the constant apex angle of the
cone (measured from the vertical axis).
Sometimes, if $\theta\in (0, \pi/2)$, the ball is said to
move in a funnel, while
if $\theta\in (\pi/2, \pi)$, the ball is said to move on a cone.
We will call the surface of rolling a {\it cone} for any $\theta$.

We will consider the motion of the ball relative to an inertial (fixed)
coordinate system $Oxyz$ in which the axis $Oz$ is
directed along the axis of rotation, that is $\bs \Omega=(0, 0, \Om)$.

Since there is no slipping, the velocity of the point of contact~$P$ on the ball coincides with
the velocity of an analogous point on the rotating surface, that is,
\begin{equation}
\label{comm_02}
\bs{v} +  R\bs{\gamma}\times \bs{\omega} = \bs{\Omega}\times\bs{r}_p, \q \bs{r}_p=\bs r-R\bs\gamma,
\end{equation}
where $\bs r=(x, y, z)$ is~the~radius vector of~the~center of mass of~the~ball,
$\bs{v}=\dot{\bs r}$ and $\bs{\omega}$ are, respectively, the velocity
of~the~center of mass and
the angular
velocity of~the~ball in~the~coordinate system $Oxyz$, $\bs{\gamma}$~is the normal at the point of
contact, $R$ is the radius of the ball, and $\bs{r}_p$ is the vector directed
from the origin of the coordinates to the point of~contact~$P$. Here and in what follows, all
vectors are written in boldface font.

The change in the linear and angular momenta of the ball relative to its center
can be written in the form of Newton--Euler equations:
\begin{equation}
m\ddot{\bs r}=\bs{N}+\bs F, \qq
I\dot{\bs\omega}=R\bs{N}\times\bs\gamma,\quad
\label{comm_03}\end{equation}
 where $m$ is the mass of the ball,  $I$ is
the central tensor of inertia of the ball, $\bs F$ is
the resultant of the
external active forces, and $\bs{N}$ is the reaction force acting on the
ball at the point of contact~$P$ (in the general case it can have any
direction).

Eliminating the reaction force from the second of Eqs.~\eqref{comm_03} and
performing a vector product by $\bs \gamma$, we obtain
\begin{equation}
\bs \gamma\times\dot{\bs\omega}=\dfrac{mR}{I}
\Big(\ddot{\bs r}-\bs \gamma(\bs \gamma \cdot \ddot{\bs r})\Big)
 -\dfrac{R}{I}\big(\bs F
-\bs \gamma(\bs \gamma \cdot \bs F)\big).
\label{comm_05}\end{equation}

The vector product on the left-hand side of \eqref{comm_05} can be
expressed from the derivative of the constraint
equation \eqref{comm_02} with respect to time:
\begin{equation}
\bs \gamma\times\dot{\bs\omega}=\dfrac{1}{R}
\Big(-\ddot{\bs r}+\bs\Omega\times\dot{\bs r}+
R\dot{\bs \gamma}\times(\bs\Omega-\bs\omega)\Big).
\label{comm_06}\end{equation}

Equating the right-hand sides of Eqs.\eqref{comm_05}
and \eqref{comm_06}, we obtain an equation
governing the evolution of the vector $\bs r$:
\begin{equation}
(k+1)\,\ddot{\bs r}-\bs\gamma \left(\bs\gamma \cdot \ddot{\bs r}\right)
= \dfrac{1}{m}\Big( \bs F-\bs\gamma \left(\bs\gamma \cdot \bs F \right)\Big)+
k\,\bs\Omega\times\dot{\bs r}+kR\dot{\bs\gamma}\times
(\bs \Omega-\bs\omega), \q k=\dfrac{I}{mR^2},
\label{comm_07}\end{equation}
where $k\geqslant0$ is the parameter of the system, $k=2/5$ for a homogeneous ball.

We see that, in order to obtain a closed system of equations, we need to define the evolution of the vectors $\bs\gamma$ and
$\bs\om$.

To do so, we first note that the vectors $\bs{r}$ and $\bs{\gamma}$ in
Eq.~\eqref{comm_07} are dependent. If the surface on which the center of
the ball moves is defined by the equation $\Phi(\bs{r})=0,$  then the
normal vector to this surface is collinear with the normal vector to the
surface on which the ball rolls, and is given by the Gauss map
\begin{equation}
\label{bmb_Eq11}
\bs{\gamma}=-\frac{\nabla \Phi(\bs{r})}{|\nabla \Phi(\bs{r}) |}.
 \end{equation}

In the case of motion of the ball on the internal surface of the cone, the coordinates of the center of the ball are related by
\begin{equation} \label{cone_Phi}
\Phi(\bs{r})=z^2\sin^2\theta -(x^2 + y^2) \cos^2\theta=0.
\end{equation}
From this equation we express the normal to~the~surface
using the~coordinates of the vector~$\bs r$ in explicit form as
\begin{equation} \label{cone_gam}
\gamma_1=-\frac{x\cos\theta}{\sqrt{x^2 + y^2}}\, , \q
\gamma_2=-\frac{y\cos\theta}{\sqrt{x^2 + y^2}}\, , \q \gamma_3=\sin\theta.
\end{equation}

Second, to determine the angular velocity $\bs\om$, we represent it in the form $\bs \om=\bs\om_\tau+\om_n\bs\gamma,$ where $\om_n=(\bs\om\cdot\bs\gamma)$ is the projection of the angular velocity onto the normal and $\bs\om_\tau$ is the projection of the angular velocity onto the plane tangent to the surface, which is determined from the constraint equation \eqref{comm_02}:
\begin{equation} \begin{gathered} \label{cone_gam02}
\bs\om_\tau=R^{-1}
\left(\bs{\gamma}\times \dot{\bs r}+(\bs\Om\times(\bs r-R\bs\gamma))\times\bs\gamma\right).
\end{gathered} \end{equation}

Thus, to close the system \eqref{comm_07}, we need to obtain the missing equation for the evolution of the projection of the angular velocity $\om_n$ onto the normal.  To find it, we perform a scalar product of the second equation of \eqref{comm_03} by $\bs \gamma$, and obtain
$k(\dot{\bs\omega}, \bs\gamma)=0,$
whence
using  $(\dot{\bs\gamma}\cdot\bs\gamma)=0$
we obtain additional equation
\begin{equation}
\label{comm_21}
\dfrac{d \omega_n}{dt}=(\bs\omega_\tau \cdot \dot{\bs\gamma}).
\end{equation}

Thus, Eqs. \eqref{comm_07} and \eqref{comm_21},
with \eqref{cone_gam}
taken into account, form a complete closed system of differential
equations governing the motion of a homogeneous ball on the surface
of~the~cone.

In this paper, we shall assume that the ball is acted upon by the
gravity force  $\bs F=m\bs g$, where $\bs g$ is the
free-fall acceleration.

\subsection{Equations of motion in spherical coordinates}
In this section, as in Ref.\cite{Zengel}, we make use of spherical
coordinates related to the Cartesian coordinates by
\begin{equation}\label{comm_12}
x=r\sin\theta\cos\varphi, \q y=r\sin\theta\sin\varphi , \q z=r\cos\theta.
\end{equation}

In the spherical coordinates, in view of Eq. \eqref{comm_12},
the surface equation \eqref{cone_Phi}
can be reduced to the following very simple form
$\theta=\const.$

Further, we write out the projections of the vectors appearing in Eqs.~\eqref{comm_02}, \eqref{comm_07},
 \eqref{comm_21}, and their time derivatives onto the spherical basis
 $\bs e_r, \bs e_\theta, \bs e_\varphi$ (see Fig. \ref{comm_01}),
 taking into account
that $\dot\theta=\ddot\theta=0$ (for details on the rules of
differentiation of vectors in spherical coordinates, see, e.g., Ref.\cite{OReil}):
\begin{equation}\label{comm_10}
\begin{gathered}
\bs r= r \bs e_r, \q \bs\gamma=- \bs e_\theta,
\q \bs\om=\om_r \bs e_r +\om_\theta \bs e_\theta + \om_\varphi \bs e_\varphi, \\
\q \dot{\bs r} =\dot r \bs e_r+ r\dot\varphi\sin\theta\bs e_\varphi,
\q \dot{\bs \gamma} =-\dot\varphi \cos\theta \bs e_\varphi,\\
\dot{\bs\omega}=(\dot\omega_r-\dot\varphi \om_\varphi \sin\theta) \bs e_r+
(\dot\omega_\theta-\dot\varphi \omega_\varphi\cos\theta) \bs e_\theta+
(\dot\omega_\varphi+\dot\varphi (\omega_\theta \cos\theta+ \omega_r\sin\theta))\bs e_\varphi,
\\
\ddot{\bs r} =(\ddot r- r\dot\varphi^2 \sin^2\theta) \bs e_r-r\dot\varphi^2\cos\theta \sin\theta \bs e_\theta+
 (r\ddot\varphi+2\dot r\dot\varphi) \sin\theta  \bs e_\varphi,\\
\bs F= -mg\cos\theta \bs e_r+ mg \sin\theta \bs e_\theta ,
\q \bs\Om= \Om\cos\theta \bs e_r -\Om\sin\theta \bs e_\theta.\\
\end{gathered}
\end{equation}

Using Eq. \eqref{comm_10}, the nontrivial
projections of Eqs. \eqref{comm_07} and \eqref{comm_21}
onto the basis $\bs e_r, \bs e_\theta, \bs e_\varphi$
can be represented in the form of five differential first-order
equations
\begin{equation}\label{comm_23}
\begin{gathered}
\dot r={V_r}, \q \dot\om_\theta=\dfrac{V_r V_\varphi}{R} \cos\theta, \q
\dot V_\varphi = \dfrac{V_r}{r} \left( \dfrac{ \Om k}{1+k}-
2 V_\varphi\right),\\
\dot V_r= r V_\varphi^2\sin^2\theta -\dfrac{k V_\varphi}{1+k}
\left(r\Om\sin^2\theta  + R \cos\theta (\Om \sin\theta
+\om_\theta) \right)-\dfrac{g\cos\theta}{1+k},
\end{gathered}
\end{equation}
\begin{equation}\label{comm_11}
\dot\varphi=V_\varphi,
\end{equation}
 and two algebraic equations
\begin{equation}\label{comm_24}
\om_r=\dfrac{r}{R}(\Om-V_\varphi)\sin\theta +\Om\cos\theta,
\q \om_\varphi=\dfrac{V_r}{R}.\\
\end{equation}

The equation for $\dot\varphi$ \eqref{comm_11}
decouples from the general system, since the variable $\varphi$
does not appear explicitly in the system \eqref{comm_23}. This is due to the invariance
of the initial system under rotations about the vertical axis.

In addition, the variables $\om_r, \om_\varphi$ do not appear
explicitly in the resulting differential equations either and
can be calculated immediately from Eq.\eqref{comm_24} after
integrating the system~\eqref{comm_23}.

Thus, the system \eqref{comm_23}
is closed relative to the unknowns
$(r, \om_\theta, V_\varphi, V_r)$ and completely defines the dynamics
of the ball on the cone. In order to restore the motion of the
center of mass of the ball in space and to construct its
trajectory in the fixed coordinate system $Oxyz$, it is
necessary to add the equation \eqref{comm_11} to the
system \eqref{comm_23} and to calculate the dependence of
the Cartesian coordinates $(x, y, z)$ on time by using their
relation to the spherical coordinates~\eqref{comm_12}.

\subsection{Comments on the paper
by K. Zengel   ``The electromagnetic analogy of a ball on a rotating conical
turntable''}
\label{comm_comm}

In the paper
by K. Zengel~\cite{Zengel},
the authors considered the case $\theta=\pi/2+\delta,$
where $\delta$ is a small angle. Using~\eqref{comm_23}, we rewrite in this case the equations
governing the evolution of $r, \varphi, \om_\theta$ in the form
\begin{equation}\label{comm_11_rem}
\begin{gathered}
0=\ddot r-r\dot\varphi^2\cos^2\delta +
\dfrac{r\Om\dot\varphi}{1+k^{-1}}\cos^2\delta
-G\dfrac{\sin\delta}{1+k},
\q  G=g+k R \,\dot\varphi (\Om\cos\delta +\om_\theta),\\
0=\dfrac{d}{dt}
\left(mr^2\cos^2\delta\left[\dot\varphi-\dfrac{k\Om}{2(1+k)}\right]\right)=\dfrac{d}{dt}\{L_z\},\\
\dot\om_\theta=-\dfrac{\dot r \dot\varphi}{R} \sin\delta,
\end{gathered}
\end{equation}
where $L_z$ is a conserved quantity of the system.

The authors~\cite{Zengel} also make the following assumption:
``the point of contact is
assumed to be located directly beneath the center of mass of
the ball. For small angles, this approximation holds, but for
steep angles and balls with large outer radii, Eq. (3) must be
modified''. Based on this, they write equations \eqref{comm_11_rem},
in which the function $G$
is chosen approximately as follows:
$G\approx g.$

We see that  in general case this approximation is incorrect,
since no account is taken of
quantities of the same order of smallness in $\delta$ as $g$.
 At the same time, we note that
the term in the function G neglected by the authors~\cite{Zengel}
\begin{equation}\label{comm_12_rem}
k R \dot\varphi(\Om\cos\delta +\om_\theta)
\end{equation}
is proportional to the radius of the ball.
In the experiments presented in \cite{Zengel},
the radius is $R=0.019$ m, and hence the neglected sum
\eqref{comm_12_rem} at the velocities $\Om, \om_\theta$ used
by the authors~\cite{Zengel} is much smaller than the value of
$g$, and it can be neglected.

\begin{figure}[!h]
\centering
\includegraphics[width=0.78\linewidth]{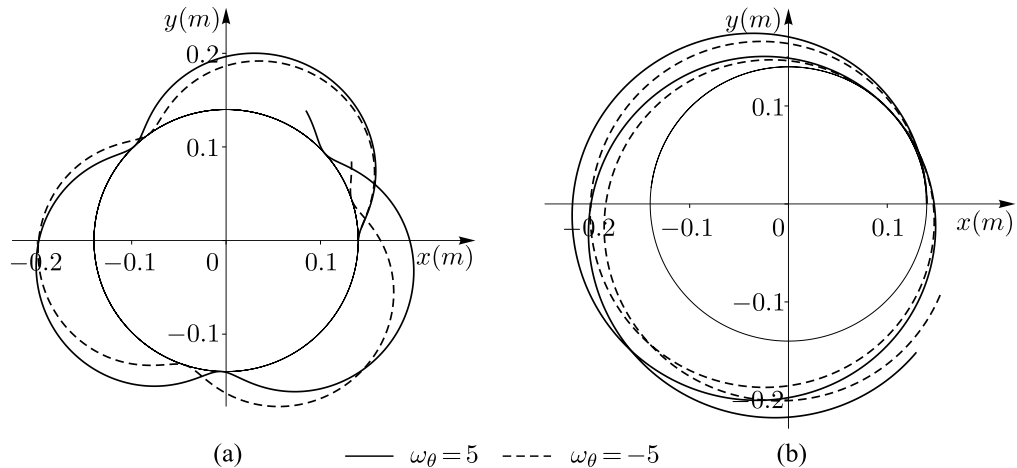}
\caption{  Projections of the trajectory of the
center of mass of the ball onto the horizontal plane
for different values of $\om_\theta$ and the parameters
$k=2/5, \Om=7$ rad/s, $g=9.8$ m/s$^2$, $R=0.2$~m, $\delta=\pi/180$
for the initial conditions (a) $\dot\varphi(0)=0.25$ rad/s
and (b) $\dot\varphi(0)=1.75$ rad/s (they correspond to
values of the quantity
$L_z/(mr_0^2\cos^2\delta)=\pm0.75$ rad/s  in \cite{Zengel}).
All trajectories begin on the right side of the plot at $x=0.14$ m
and $y=0$ m. For a comparison, the thin solid line denotes the corresponding periodic
trajectories.}
\label{comm_Tr}
\end{figure}

As the radius of the ball increases (the other
parameters and initial conditions being equal) the trajectory
becomes sensitive to the initial value of the angular velocity
$\omega_\theta$
(see Fig. \ref{comm_Tr}). In this case,
the equations governing the evolution of the angular
velocity $\om_\theta$ and the equations for the evolution of $r$ and $\varphi$ do not decouple and must be
solved jointly.

In particular, this leads to the result that, for a cone,
it is impossible to write the corresponding equations of motion
in a natural Hamiltonian form, because the phase space dimension is 5.
We note that the problem of Hamiltonization of nonholonomic
systems is fairly
complicated. Discussions on this topic can be found
in \cite{OhsawaBloch, Koiller, BolBM_Ham}.

\section{Periodic solutions}

In this section we investigate the circular orbits,
i.e., the motion of the center of
mass of the ball on a cone in the horizontal plane with some
constant frequency.

In the case of the circular periodic motion of the center
of mass, the following relations must be satisfied:
\begin{equation}\label{comm_14}
\begin{gathered}
r=r_0=\const, \q V_r=0,   \q V_\varphi=\om_0=\const,
\q \dot V_r=0,  \q \dot V_\varphi=0,
\end{gathered}
\end{equation}
and $r_0>0$.

From the equation for $\dot \om_\theta$ \eqref{comm_23} we obtain
\begin{equation}\label{comm_15}
\dot\om_\theta=0, \q \om_\theta=\Om_\theta=\const.
\end{equation}

Consequently, in the case of motion in circular orbits
according to Eqs. \eqref{comm_14},
\eqref{comm_15} and the first of Eqs. \eqref{comm_23}, the
derivatives $\dot r, \dot\om_\theta, \dot V_r, \dot V_\varphi$
vanish, which corresponds to definition of the fixed points
of the reduced system
 \eqref{comm_23}
 or periodic motions of the complete system~\eqref{comm_23}--\eqref{comm_11}.

The constants $r_0, \om_0, \Om_\theta$  parameterize
the circular orbits under consideration. Substituting Eqs. \eqref{comm_14} and \eqref{comm_15}
into the third of Eqs. \eqref{comm_23}, we obtain an equation
that relates these parameters:
\begin{equation}\label{comm_19}
 \om_0^2 -\dfrac{k \left(r_0\Om\sin^2\theta+
R \cos\theta(\Om\sin\theta+\Om_\theta)
 \right)}{r_0(1+k)\sin^2\theta}
 \om_0-\dfrac{g\cos\theta}{r_0(1+k)\sin^2\theta}=0.
\end{equation}

Thus, only two parameters are independent.
Following \cite{Zengel}, as independent parameters we choose
$r_0$ and $\Om_\theta$, and find the third parameter, $\om_0$,  from equation
\eqref{comm_19} (quadratic in $\om_0$).

Eq. \eqref{comm_19} can have two, one or no root $\om_0$,
depending on the sign of the discriminant, which has the form
\begin{equation}\label{comm_13}
\begin{gathered}
D(r_0, \Om_\theta)= \dfrac{k^2
\left(r_0\Om\sin^2\theta+
R \cos\theta(\Om\sin\theta+\Om_\theta)
 \right)^2}{r_0^2(1+k)^2\sin^4\theta}+
\dfrac{4g\cos\theta}{r_0(1+k)\sin^2\theta}.
\end{gathered}
\end{equation}

Let us analyze in more detail the solution \eqref{comm_19}
in all possible
cases for different $\theta$.

\textbf{1. } When $\theta\in(0,\pi/2)$, the discriminant \eqref{comm_13} is positive
for all possible values of the system parameters.
Consequently,\textit{ for any values of $r_0>0$, $\Om_\theta$ and
$\theta\in(0,\pi/2)$
there exist two families of periodic solutions to the system
\eqref{comm_23}-\eqref{comm_11} which correspond to different frequencies of stationary motion
}
\begin{equation}\label{comm_17}
\begin{gathered}
\om_{01}(r_0, \Om_\theta)=\dfrac{B(r_0, \Om_\theta)}{2}
+\dfrac{\sqrt{D(r_0, \Om_\theta)}}{2},
\q \om_{02}(r_0, \Om_\theta)=\dfrac{B(r_0, \Om_\theta)}{2}-
\dfrac{\sqrt{D(r_0, \Om_\theta)}}{2},
\end{gathered}
\end{equation}
where $B(r_0, \Om_\theta)$ has been introduced to abbreviate the formula:
\begin{equation}\label{comm_BB}
B(r_0, \Om_\theta)=\dfrac{k
\left(r_0\Om\sin^2\theta+
R \cos\theta(\Om\sin\theta+\Om_\theta)
 \right)}{(1+k)r_0\sin^2\theta}.
\end{equation}

According to \eqref{comm_17}, the resulting frequencies
 correspond to motion of the center of mass of the ball in a circle in opposite directions.

\textbf{2. }When $\theta=\pi/2$ (horizontal plane), equation
 \eqref{comm_19} is transformed to an equation linear in $\om_0$, from which
we obtain a well-known expression for the frequency of the circular motion of the ball
on the rotating plane in the case of nonholonomic rolling (see \cite{Zengel, cone_BIKM_2018}
and references therein):
$$\om_0=\dfrac{k\Om}{k+1}.$$

\textbf{3. }\label{comm_pro03} When $\theta\in(\pi/2, \pi)$ (which is the case considered
in detail in \cite{Zengel}), given $(r_0, \Om_\theta)$, the discriminant \eqref{comm_13}
can take both positive and negative values.
Let us represent it as
\begin{equation}\label{comm_135}
\begin{gathered}
D(r_0, \Om_\theta)=\dfrac{\Om^2k^2}{r_0^2(1+k)^2}
(r_0^2+a_1 r_0+a_2),
\q \theta=\pi/2+\delta, \q \delta\in(0, \pi/2),\\
a_1=-\dfrac{2\sin\delta}{k^2\Om^2\cos^2\delta}
( Rk^2 \Om(\Om\cos\delta+\Om_\theta)+2g(1+k)),\q
a_2=\dfrac{R^2\sin^2\delta}{\Om^2\cos^4\delta}(\Om\cos\delta+\Om_\theta)^2.
\end{gathered}
\end{equation}

According to \eqref{comm_135}, the sign of $D(r_0, \Om_\theta)$ is defined
by the polynomial $r_0^2+a_1 r_0+ a_2$ quadratic in $r_0$. As is well known, on the interval
$r_0>0$
this polynomial either has no roots and is positive for all $r_0$ or has two roots
\begin{equation}\label{comm_138}r_{-}(\Om_\theta)=-\dfrac{a_1}{2}-\dfrac{\sqrt{a_1^2-4a_2}}{2},
\q r_{+}(\Om_\theta)=-\dfrac{a_1}{2}+\dfrac{\sqrt{a_1^2-4a_2}}{2},
\end{equation}
and on the interval $r_0\in(r_{-}(\Om_\theta), r_{+}(\Om_\theta))$ this polynomial is negative.

The roots
$r_{-}(\Om_\theta)$  and $ r_{+}(\Om_\theta)$ exist if the inequalities
$a_1^2-4a_2>0$ and $a_1<0$ are satisfied simultaneously.
Substituting $a_1, a_2$ from \eqref{comm_135} into these inequalities gives
\begin{equation}\label{comm_136}
\Om (\Om\cos\delta+\Om_\theta)>-\dfrac{g(k+1)}{R k^2},
\end{equation}
\begin{equation}\label{comm_137}
\Om (\Om\cos\delta+\Om_\theta)>-2\dfrac{g(k+1)}{R k^2}.
\end{equation}

We see that inequality \eqref{comm_137} follows from inequality \eqref{comm_136}
and can be omitted.

Thus, we conclude that
\textit{if at $\delta\in(0, \pi/2)$
the projection of the angular velocity of the ball onto the normal $\Om_\theta$ satisfies
inequality \eqref{comm_136}, then there exist no periodic solutions to the system
\eqref{comm_23}-\eqref{comm_11} on the interval
$r_0\in(r_{-}(\Om_\theta), r_{+}(\Om_\theta))$}.

Otherwise periodic solutions exist for all $r_0\in(0, +\infty)$. Moreover,
as in the case $\theta\in(0, \pi/2)$, there exist two different
frequencies of periodic motion~\eqref{comm_17}.

\begin{rem}
We see that at small $\delta$ the coefficients of the polynomial in
\eqref{comm_135} are ${a_1\sim \delta}$, ${a_2\sim\delta^2}$. It turns out that
$r_{-}(\Om_\theta)\sim\delta$ and hence the interval $(0, r_{-}(\Om_\theta))$
is small and is almost unobservable at physical values of the system parameters.
That is, the conclusion~\cite{Zengel} that there is a minimal value
$r_{min}$ for $r_0$ on~the~cone at~$\theta\in(\pi/2, \pi)$
is~justified from a~physical point of~view.
On the other hand, in our case, $r_{min}=r_{+}(\Om_\theta)$,
and this expression also differs from $r_{min}$ found in
\cite{Zengel}. This difference is also a consequence of the incorrectness (discussed above)
of the approximation~\cite{Zengel}.
\end{rem}

\section{Stability analysis of periodic solutions}

 Let us analyze the stability of the periodic
solutions
of the system~\eqref{comm_23}--\eqref{comm_11},
which correspond to fixed points of the reduced system~\eqref{comm_23}.
For this, we linearize the system \eqref{comm_23}
near the partial solution of \eqref{comm_14}--\eqref{comm_15}.

The linearized system can be represented as
$\dot{\tilde{{\bs z}}}={\bf L}\tilde{\bs z},$
where ${\bf L}$ is the linearization matrix,
$\tilde{\bs z}=\bs z-\bs z_0$,
$\bs z=( r,  \omega_\theta, V_r, V_\varphi)$ is the vector of
the variables, and $\bs z_0= (r_0,  \Om_\theta, 0, \om_0)$ is the value of the
vectors of the variables which corresponds to
the partial solution of \eqref{comm_14}--\eqref{comm_15}.

The characteristic equation $\det({\bf L}-\lambda{\bf E})=0$
(with $\lm$ being its roots)
is
\begin{equation}
\label{comm_18}
\lm^2 \left(\lm^2+\dfrac{(k+\sin^2\theta)\om_0^2}{1+k}+\dfrac{2g\cos\theta (\om_0-\om_0^*)}{r_0 \om_0(1+k)}\right)=0,
\q \om_0^*=\dfrac{k\Om}{2(1+k)}.
\end{equation}

  In this case, to facilitate calculations,
we have chosen $r_0$
and $\om_0$ as independent parameters, and $\Om_\theta$ is
uniquely expressed from~\eqref{comm_19}.

Two zero roots of the characteristic equation correspond to the
 parameters $r_0, \om_0$ of the family of periodic solution
of the system \eqref{comm_23}-\eqref{comm_11}.
The nonzero roots of Eq.\eqref{comm_18} can be imaginary
or real (of different signs) depending on the
parameters $r_0$ and $\om_0$
and the value of the angular velocity of rotation of the cone $\Om$.

Let is consider all possible cases for different $\theta$ and  $\Om\geqslant0$.
\vspace{-4pt}
\begin{enumerate}\itemsep=-2pt
\item  When $\theta\in(0, \pi/2)$ and $\Om>0$,
the~nonzero roots of Eq. \eqref{comm_18} are imaginary,
the~periodic solutions of the system~\eqref{comm_23}-\eqref{comm_11}
are linearly stable and of center type in the cases
   \begin{enumerate}\itemsep=-3pt
    \item $\om_0>\om_0^*$ or  $\om_0<0$ for any $r_0$,
    \item $0<\om_0<\om_0^*$ and
$r_0>\dfrac{2g\cos\theta(\om_0^*-\om_0)}{\om_0^3(k+\sin^2\theta)}$.
    \end{enumerate}

\item  When $\theta\in(\pi/2, \pi)$ and $\Om>0$,
the nonzero roots of Eq.\eqref{comm_18} are imaginary,
the periodic solutions of the system~\eqref{comm_23}-\eqref{comm_11}
are linearly stable and of center type in the cases
   \begin{enumerate}\itemsep=-3pt
    \item $0<\om_0<\om_0^*$ for any $r_0$,
    \item $\om_0>\om_0^*$ or  $\om_0<0$ and
$r_0>\dfrac{2g\sin\delta(\om_0-\om_0^*)}{\om_0^3(k+\cos^2\delta)},$ where
$\delta=\theta-\pi/2$.
    \end{enumerate}

\item When $\Om=0$ and any $\theta\in(0, \pi)$, the nonzero roots of Eq.\eqref{comm_18}
are imaginary, the family of
periodic solutions of the system~\eqref{comm_23}--\eqref{comm_11}
is linearly stable
and of center type for any $\om_0\not=0$ and $r_0.$ When $\om_0=0$, according to
Eq. \eqref{comm_19}, there exist no periodic solutions
of~the~system \eqref{comm_23}--\eqref{comm_11}.
For the stability and values of the frequencies of~periodic
motions at $\Om=0$, including the case where $k=0$, see also \cite{cone_Gary}.
\end{enumerate}

\begin{figure}[!h]
\centering
\includegraphics[width=0.85\linewidth]{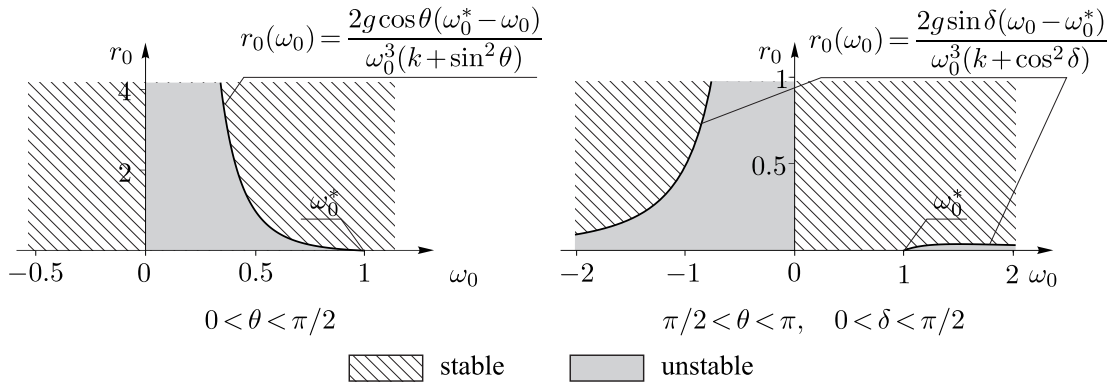}
\caption{Stability and instability regions
of the periodic solution of the system~\eqref{comm_23}--\eqref{comm_11}
on the plane $(r_0,\om_0)$ are  constructed for the parameters
$k=2/5$, $\Om=7$ rad/s, $g=9.8$ m/s$^2$  and (a)~$\theta=89\pi/180$, (b) $\theta=91\pi/180$, }
\label{comm_20}
\end{figure}

 In other cases, the family of periodic solutions of the
system~\eqref{comm_23}--\eqref{comm_11} is linearly unstable
and of saddle type. For the sake of illustration,
 stability and instability regions for $\theta\in(0, \pi/2)$ and
$\theta\in(\pi/2, \pi)$
are shown in grey in Fig. \ref{comm_20} on the plane $(\om_0, r_0)$.

We note that that the characteristic polynomial \eqref{comm_18} is
invariant under the substitution $(\Om,\om_0, r_0)\rightarrow
(-\Om,-\om_0, r_0)$.  Thus, for  $\Om<0$   analogous regions of stability
(instability) of  periodic solutions of the system~\eqref{comm_23},
\eqref{comm_11}  are obtained by a symmetric mapping of  regions of
stability (instability)   in Fig. \ref{comm_20} relative to the vertical
axis $\om_0=0$.

\section{Conclusion}

The analysis of motion of the homogeneous ball on the cone is not restricted to investigating
a partial periodic solution. The problem that remains open is that of exploring the rolling
of the ball on the cone depending on initial conditions in the general case.

Another open problem is to examine the rolling of the ball with rolling resistance.
In the recent paper \cite{cone_BIKM_2018}, in the case of motion
of a homogeneous ball on a plane, a good agreement was shown between the experimental trajectory
and the theoretical trajectory obtained by adding the moment of rolling friction which is
proportional to the angular velocity
of the ball. Using this friction model, it was shown that all
trajectories asymptotically tend to an untwisting spiral.

The preliminary numerical experiment has shown that
in the case of motion on a cone with $\theta\in(0, \pi/2)$,
as opposed to motion on a plane,
ball can  either move in an untwisting trajectory
(the value of $\rho$ and the height increase in this case)
or approach the vertex of the cone (the value of $\rho$ and the height decrease).

\section*{Acknowledgments}
The work of I.\,S.\,Mamaev and T.\,B.\,Ivanova  (Section 1) was carried out within the framework
of the state assignment of the Ministry of Education and Science of Russia (1.2405.2017/4.6).
The work of A.\,V.\,Borisov and A.\,A.\,Kilin (Sections  2, 3) was carried out at MIPT within the framework
of the Project 5-100 for State Support for Leading Universities of the Russian Federation.

\end{document}